\documentstyle[12pt,graphicx]{article}
\newcommand{\nd}{{\cal D}\hspace{-1.5ex}/}
\newcommand{\arcsinh}{\rm arcsinh}
\newcommand{\arctanh}{\rm arctanh}
\newcommand{\cossech}{\rm cossech}
\begin{document}
\title{Vacuum Polarization for a Massless Spin-$1/2$ Field in
the Global Monopole Spacetime at Nonzero Temperature}
\author{Vacuum Polarization for a Massless Scalar Field in the Global 
Monopole Spacetime at Finite Temperature}
\author{F. C. Carvalho \\ and 
E. R. Bezerra de Mello \thanks{E-mail: emello@fisica.ufpb.br}\\
Departamento de F\'{\i}sica-CCEN\\
Universidade Federal da Para\'{\i}ba\\
58.059-970, J. Pessoa, PB\\
C. Postal 5.008\\
Brazil}
\maketitle
\begin{abstract}
In this paper we present the effects produced by the temperature in the
renormalized vacuum expectation value of the zero-zero component of the
energy-momentum tensor associated with massless left-handed spinor field
in the pointlike global monopole spacetime. In order to develop this 
calculation we had to obtain the Euclidean thermal Green function
in this background. Because the expression obtained for the thermal energy 
density cannot be expressed in a closed form, its explicit dependence on the 
temperature is not completely evident. So, in order to obtain concrete 
information about its thermal behavior, we develop a numerical analysis of 
our result in the high-temperature limit for specific values of the parameter 
$\alpha$ which codify the presence of the monopole. 
\end{abstract}

\newpage
\renewcommand{\thesection}{\arabic{section}.}
\section{Introduction}

The Euclidean Green function associated with a massless left-handed spin-
1/2 field in the pointlike global monopole spacetime, $S_F(x,x')$, was 
presented recently in Ref. \cite{Mello1}. There, $S_F(x,x')$ was obtained 
applying the Dirac operator $\nd_L$, on the two-components bispinor Green
function $G_F(x,x')$, which obeys a non-homogeneous second order 
differential equation similar to the scalar case. Using this procedure 
we can write
\begin{equation}
S_F(x,x')=\nd_LG_F(x,x').
\end{equation}

The massless left-handed spinor thermal Green function, $S_{\beta
F}(x,x')$, can also be obtained from the respective thermal bispinor
$G_{\beta F}(x,x')$ in similar way. This fermionic Green function
should be antiperiodic in the ``Euclidean'' time $\tau$ with period 
$\beta=1/\kappa_B T$, $\kappa_B$ being the Boltzman constant and $T$ the 
absolute temperature.

The gravitational effect produced by a global monopole can be approximately
described by a solid angle deficit in the $(3+1)$-dimensional spacetime whose
line element is given by \cite{BV}
\begin{equation}
\label{G-M}
ds^2=-dt^2+\frac{dr^2}{\alpha^2}+r^2 (d\theta^2+\sin^2\theta d\phi^2)\ , 
\end{equation}
where the parameter $\alpha$, which codify the presence of this object is 
smaller than unity. The energy-momentum tensor associated with this monopole
has a diagonal form and reads: $T^0_0=T^1_1=(\alpha^2-1)/r^2$ and $T^2_3=
T^3_3=0$. 

Because we are interested in obtaining the thermal Green function, it is
convenient to work in the Euclidean continuation of the Green function by
performing a Wick rotation. So we shall work on the Euclidean version of
the monopole metric above, expressed by the line element
\begin{equation}
\label{B-V1}
ds^2=d\tau^2+\frac{dr^2}{\alpha^2}+r^2 (d\theta^2+\sin^2\theta d\phi^2)\ .
\end{equation}

This paper is organized as follows. In section 2 we derive the Euclidean 
thermal Green function associated with a massless left-handed spin-1/2
field in the pointlike global monopole spacetime adopting the 
imaginary-time approach \cite{Jackiw} and using the Schwinger-De Witt 
formalism. In section 3 we present a formal expression for the thermal 
average of the zero-zero component of the energy-momentum tensor, 
$\langle T_{00}(x)\rangle_\beta$. Because this term cannot be written 
in a closed form, its dependence on the temperature is not evident. So, 
in order to obtain some quantitative information about its behavior we 
derive more workable expressions for specific values of the 
parameter $\alpha$. Moreover, the complete information about its 
dependence on the temperature requires that we proceed  a numerical 
evaluation of some non-analytical contributions. Here in this paper, 
we develop these analysis in the  high-temperature regime. In section 4 
we summarize our results and give our conclusion about the thermal 
bath of massless fermions in the global monopole spacetime. In the 
Appendix we derive explicitly the thermal average for the energy
density for the special cases where $\alpha=1/2$ and give the dominant
thermal contributions for the cases $\alpha=1/3$ and $\alpha=1/4$.

\section{Thermal Spinor Green Function}

The Green function associated with a massless left-handed spin-1/2 field
in the global monopole spacetime has been obtained recently \cite{Mello1}.
There this Green function is presented by applying the respective
Dirac operator on the two-components bispinor as we shall summarize below. 

In order to write down the Dirac operator in the spacetime described by 
(\ref{G-M}) we choose the following basis tetrad:
\begin{equation}
e^\mu_{(a)} = \left( \begin{array}{cccc}
  1 & 0 & 0 & 0 \\
  0 & \alpha\sin\theta \cos\varphi & \cos \theta
  \cos\varphi /r & -\sin\varphi /r\sin\theta \\ 
0 & \alpha\sin\theta \sin\varphi & \cos \theta
\sin\varphi /r & \cos\varphi /r\sin\theta \\ 
0 & \alpha\cos\theta & -\sin \theta /r & 0 
                      \end{array}
               \right) \ .
\end{equation}
With this choice the Dirac operator for a massless left-handed fermionic 
field in a global monopole spacetime given by
\begin{equation}
\nd_L=e^\mu_{(a)}\sigma^{(a)}\left(\partial_\mu+\Gamma_\mu\right)\ ,
\end{equation}
reduces itself to\footnote{Here $\sigma^{(a)}=(I,\sigma^k)$, $\sigma^k$, 
with $k=1,2,3$, being the $2\times2$ Pauli matrices and $\Gamma_\mu$ the 
spin connection.}
\begin{equation}
\label{Dirac}
\nd_L=i\left(\partial_t-\alpha\sigma^r\partial_r-\frac 1 r \sigma^\theta
\partial_\theta-\frac1{r\sin\theta}\sigma^\phi\partial_\phi+\frac{1-\alpha}r
\sigma^r\right) \ ,
\end{equation}
with $\sigma^u=\vec{\sigma}\cdot\hat{\vec{u}}$, where $\hat{\vec{u}}$ denotes 
the standard unit vector along the three spatial directions in spherical
coordinates.

The Feynman two-components propagator obeys the differential equation
\begin{equation}
\nd_L S_F(x,x')=\frac 1{\sqrt{-g}}\delta^{(4)}(x-x')I_{(2)}
\end{equation}
and can be given in terms of the bispinor $G_F(x,x')$ by
\begin{equation}
\label{Spinor}
S_F(x,x')=i\nd_LG_F(x,x') \ ,
\end{equation}
where $G_F(x,x')$ obeys the $2\times2$ differential equation
\begin{equation}
{\cal{L}}G_F(x,x')=-\frac 1{\sqrt{-g}}\delta^{(4)}(x-x')I_{(2)}\ ,
\end{equation}
with
\begin{equation}
{\cal{L}}=\Box-\frac1 4 R\ .
\end{equation}
In the above equation, the generalized d'Alembertian operator is 
expressed by
\begin{equation}
\Box=g^{\mu\nu}\nabla_\mu\nabla_\nu=g^{\mu\nu}\left(\partial_\mu
\nabla_\nu+\Gamma_\mu\nabla_\nu-\Gamma^\alpha_{\mu\alpha}\nabla_\alpha
\right) \ ,
\end{equation}
and $R$, the scalar curvature, is equal to $2(1-\alpha^2)/r^2$ for this
spacetime.

From now on we shall work with the Euclidean version of the spinor
Green function. The Euclidean bispinor, $G_E(x,x')$, can also be obtained 
from the Schwinger-DeWitt formalism as follows:
\begin{equation}
\label{G_infty}
G_(x,x)=\int_0^\infty ds K(x,x';s)\ ,
\end{equation}
where the heat kernel, $K(x,x';s)$, can be expressed in terms of the
eigenfunctions of the operator $\cal{L}$ as
\begin{equation}
K(x,x';s)=\sum_\lambda \Phi_\lambda(x)\Phi_\lambda^\dagger(x')e^{-s\lambda^2}\ ,
\label{K}
\end{equation}
$\lambda^2$ being the corresponding positively defined  eigenvalue.
Writing
\begin{equation}
{\cal{L}}\Phi(x)=-\lambda^2\Phi(x) \ ,
\end{equation}
the complete set of normalized solutions of the above equation is given 
by \cite{Mello1}\footnote{In Ref. \cite{Mello1} we have used another 
parametrization for the time component. There $g_{00}=\alpha^2$. For this 
reason there is a small changing in the expressions obtained for the 
normalization of the field itself and its eigenvalue.}:
\begin{eqnarray}
\Phi^{(k)}_\lambda(x) &=& \sqrt{\alpha p \over 2\pi r} e^{-iE\tau}
J_{\nu_k} (pr) \varphi^{(k)}_{j,m_j}\ , \label{Phi}\\
\lambda^2 &=& E^2+ \alpha^2 p^2\ , \nonumber \\
\nu_1 &=& {l+1 \over \alpha} - {1\over 2}\ ,\ \nu_2 = {l\over
\alpha} + {1 \over 2}\ , \label{nu}
\end{eqnarray}
where $J_\nu$ is the Bessel function of the first kind and $\varphi^{(k)}_{j,m_j}$,
with $k=1, 2$, are the spinor spherical harmonics eigenfunctions of the operators
${\vec L}^2$ and $\sigma\cdot\vec L$. (The explicit expressions for these spherical 
harmonics and their eigenvalue equations are given in \cite{BD}. There, it is explicitly
shown the dependence of $j$ with $l$.)

Now substituting (\ref{Phi}) into (\ref{K}) we get:
\begin{eqnarray}
\label{K_infty}
K_{\infty}(x,x';s)&=&\int_{-\infty}^{\infty}\frac{d\omega}{2\pi}
\int_{0}^{\infty}dp\sum_{l,m}\Phi_{\lambda}(x)\Phi^\dagger_{\lambda}(x')
e^{-s\lambda^2}
\nonumber\\
&=&\frac 1{4\alpha\sqrt{\pi rr'}}\frac1{s^{3/2}}\exp^{-\frac{(\Delta
\tau)^2\alpha^2+r^2+r'^2}{4s\alpha^2}}\sum_{j,m_j}\left[I_{\nu_1}
\left(\frac{rr'}{2\alpha^2 s}\right)C_{j,m_j}^{(1)}(\Omega,\Omega') \right.
\nonumber\\
&&+\left.I_{\nu_2}\left(\frac{rr'}{2\alpha^2 s}\right)
C_{j,m_j}^{(2)}(\Omega,\Omega')\right]  \ ,
\nonumber\\
\end{eqnarray}
where $I_\nu$ is the modified Bessel function and
$C^{(k)}_{j,m_j}(\Omega ,\Omega')=\varphi^{(k)}_{j,m_j}(\Omega) 
\varphi^{(k)\dagger}_{j,m_j}(\Omega')$. With the help
of \cite{G} we can see that for $\alpha=1$, $\nu_1=\nu_2=l+1/2$, it is possible
to proceed the two summations on the quantum numbers $j$ and $m_j$ and (\ref{K_infty})
reduces itself to:
\begin{equation}
K_\infty^{(\alpha=1)}(x,x';s)=\frac 1{16\pi^2}\frac{e^{-\frac{(x-x')^2}{4s}}}{s^2}I_{(2)}\ .
\end{equation}

Now we are in position to obtain the Euclidean bispinor $G_E(x,x')$ by using
(\ref{G_infty}). Again with the help of \cite{G} we obtain:
\begin{eqnarray}
\label{GreenE}
G_E(x,x') = {1 \over 2\pi r r'} \sum_{j,m_j} \left[
Q_{\nu_1 - 1/2} (u) C^{(1)}_{j,m_j}(\Omega ,\Omega')\right. + \nonumber\\
\left. Q_{\nu_2
- 1/2} (u) C^{(2)}_{j,m_j}(\Omega , \Omega')\right]\ , 
\end{eqnarray}
$Q_\mu$ being the Legendre function and
\begin{equation}
u=\frac{\alpha^2(\tau-\tau')^2+r^2+r'^2}{2rr'}\ .
\end{equation}

The Euclidean thermal bispinor can be obtained in the same way by using
the thermal heat kernel, $K_\beta(x,x';s)$:
\begin{equation}
\label{Heat_beta}
G_\beta(x,x')=\int_0^\infty ds K_\beta(x,x';s) \ .
\end{equation}

For an ultrastatic spacetime, i.e., static with $g_{00}=1$, Braden 
\cite{Braden} has proved that the thermal heat kernel can be expressed in 
terms of the sum
\begin{equation}
K_\beta(x,x';s)=\sum_{n=-\infty}^\infty (-1)^n K_\infty(x,x'-n\lambda\beta;s) \ ,
\end{equation}
where $\lambda$ is the ``Euclidean'' time unit vector. The zero-temperature
heat kernel, $K_\infty(x,x';s)$, given previously by (\ref{K_infty}) can be
factorized as:
\begin{equation}
K_\infty(x,x';s)=K_{(1)}(\tau,\tau';s)K_{(3)}(\vec x, \vec x';s) \ ,
\end{equation}
where $K_{(1)}(\tau,\tau';s)$ and $K_{(3)}(\vec x,\vec x';s)$ obey, respectively,
the differential equations below:
\begin{equation}
\label{K_1}
\left(\frac{\partial}{\partial s}-\frac{\partial^2}{\partial \tau^2}\right)
K_{(1)}(\tau,\tau';s)=0
\end{equation}
and
\begin{equation}
\left(\frac{\partial}{\partial s}-\nabla^i\nabla_i+\frac14 R\right)
K_{(3)}(\vec x,\vec x';s)=0\ .
\end{equation}

Identifying in (\ref{K_infty}) as the solution for (\ref{K_1})
\begin{equation}
K_{(1)}(\tau,\tau';s)=\frac{e^{-\frac{\Delta\tau^2}{4s}}}{\sqrt s}\ ,
\end{equation}
we can see that only $K_{(1)}(\tau,\tau';s)$ will be affected by the
temperature, so we obtain
\begin{equation}
K_{(1)\beta}(\tau,\tau';s)=\frac1{\sqrt s}\sum_{n=-\infty}^\infty(-1)^n
e^{-\frac{(\Delta\tau+n\beta)^2}{4s}} \ ,
\end{equation}
which is antiperiodic in coordinate $\tau$. Now we can write down our
result for the thermal heat kernel:
\begin{eqnarray}
\label{K_beta}
K_\beta(x,x';s)&=&\frac 1{4\alpha\sqrt{\pi rr'}}\frac 1 {s^{3/2}}
\sum_{n=-\infty}^\infty(-1)^n\exp^{-\frac{(\Delta\tau+n\beta)^2\alpha^2+r^2+r'^2}
{4s\alpha^2}}
\nonumber\\
&&\sum_{j,m_j}\left[I_{\nu_1}\left(\frac{rr'}{2\alpha^2 s}\right)
C_{j,m_j}^{(1)}(\Omega,\Omega')+I_{\nu_2}\left(\frac{rr'}{2\alpha^2 s}\right)
C_{j,m_j}^{(2)}(\Omega,\Omega')\right]  \ .
\nonumber\\
\end{eqnarray}

Finally substituting (\ref{K_beta}) into (\ref{Heat_beta}), we obtain an
explicit expression for the thermal bispinor, $G_\beta(x,x')$:
\begin{eqnarray}
\label{Green}
G_\beta(x,x')&=&\frac1{2\pi rr'}\sum_{n=-\infty}^\infty(-1)^n\sum_{j,m_j}\left[
Q_{\nu_1-1/2}(u_{n\beta})C_{j,m_j}^{(1)}(\Omega,\Omega')\right.
\nonumber\\
&&\left.+Q_{\nu_2-1/2}(u_{n\beta})C_{j,m_j}^{(2)}(\Omega,\Omega')\right] \ ,
\end{eqnarray}
where
\begin{equation}
u_{n\beta}=\frac{\alpha^2(\Delta\tau+n\beta)^2+r^2+r'^2}{2rr'} \ .
\end{equation}

As we can see form (\ref{Green}), the thermal bispinor contains explicitly
the zero-temperature Green function, $G_E(x,x')$, plus purely thermal 
contributions. Moreover, because the Legendre functions, $Q_\mu(z)$, decrease 
proportionally with $1/z^{\mu+1}$ for large arguments \cite{G}, we conclude 
that, for a given set of values of the quantum numbers $j$ and $m_j$,
the most relevant thermal corrections to $G_\beta(x,x')$ above, come from 
the first terms in the series in $n$.

Finally we can write down our Euclidean thermal spinor Green function
substituting (\ref{Green}) and the Euclidean version of (\ref{Dirac}) 
into (\ref{Spinor}):
\begin{equation}
\label{S_beta}
S_\beta(x,x')=i\left(-i\partial_\tau-\alpha\sigma^{(r)}\partial_r+
 \frac 1 r\sigma^{(r)}\vec{\sigma}\cdot\vec L +\frac{1-\alpha}r
\sigma^r\right)G_\beta(x,x')  \ .\
\end{equation}

\section{Thermal Average of the Energy-Momentum Tensor}

The energy-momentum tensor associated with spin-$1/2$ fields, $T_{\mu\nu}(x)$,
is bilinear function of the fields. Consequently its vacuum expectation 
value (VEV), $\langle T_{\mu\nu}(x)\rangle$, can be evaluated by the
standard method using the Green function \cite{Birrel}.

In \cite{Mello1} the renormalized VEV of the energy-momentum tensor
associated with a left-handed spin-$1/2$ field in a global monopole
spacetime was explicitly calculated. There, it was shown that it has
the following structure:
\begin{equation}
\label{T-mu,nu}
\langle T^\nu_\mu(x)\rangle^{Reg.}=\frac 1{8\pi^2r^4}\left[A^\nu_\mu+
B^\nu_\mu ln(\mu r)\right] \ ,
\end{equation}
where the scaling parameter $\mu$ appears after the renormalization
procedure. In fact, applying the point-splitting renormalization
procedure and extracting from the spinor Green function all the
divergences by subtracting the correspondent Hadamard function,
the renormalized VEV of the energy-momentum tensor presents a
scale-dependent logarithmic term. This is a consequence of an
ambiguity in the definition of $\langle T_{\mu\nu}\rangle^{Reg.}$ in a
even dimensional curved spacetime. The back-reaction probelm consists
of solving the Einstein equaion in presence of quantum corrections.
So, this one-loop equation contains besides $\langle T_{\mu\nu}
\rangle^{Reg.}$ as source of energy-momentum tensor, extra purely
geometrical terms \cite{Birrel}. The coefficients of these terms
must depend on the renormalization scale since the theory must
be independent of this scale due to renormalization group equation. 
The tensors $A^\nu_\mu$ and $B^\nu_\mu$ depend only on the 
metric parameter $\alpha$. Moreover, because of the spherical
symmetry these tensors are diagonal with components $A^\theta_\theta=
A^\phi_\phi$ and $B^\theta_\theta=B^\phi_\phi$. Therefore there are only
six unknown components. The renormalized VEV of the energy-momentum tensor
must be conserved, i.e.,
\begin{equation}
\label{Conservation}
\langle T^\nu_\mu \rangle^{Reg.}_{;\nu}=0 \ ,
\end{equation}
and gives the correct conformal anomaly \cite{Wald}, which for massless
spinor two-components field, reads \cite{CD}
\begin{equation}
\label{Anomaly}
\langle T^\mu_\mu\rangle ^{Reg.}=\frac1{16\pi^2}tr a_2=
\frac {\cal{T}}{8\pi^2r^4}\ .
\end{equation}
Here we have introduced the new definition ${\cal{T}}:=\frac{r^4tr a_2}2$. The
general form of the coefficient $a_2$ may be found in \cite{Christensen},
which for the global monopole spacetime has the following form:
\begin{equation}
a_2=-\frac{1-\alpha^4}{60r^4}I_{(2)}\ .
\end{equation}

Taking into account (\ref{Conservation}) and (\ref{Anomaly}) it is
possible to express the tensors $A^\nu_\mu$ and $B^\nu_\mu$ in terms of
their zero-zero components and the trace $T$ by
\begin{equation}
A^\nu_\mu=diag\left(A^0_0,-{\cal{T}}+A^0_0+B^0_0,{\cal{T}}-A^0_0-\frac{B^0_0}2,
{\cal{T}}-A^0_0-\frac{B^0_0}2\right) \ ,
\end{equation}
and
\begin{equation}
B^\nu_\mu=B^0_0 diag (1,1,-1,-1)\ .
\end{equation}
So the only problem left is to determine the components $A^0_0$ and 
$B^0_0$. Using the point-splitting approach, the VEV of the energy-momentum
tensor for a massless spinor field is given by:
\begin{equation}
\label{T_mu,nu}
\langle T_{\mu\nu}(x)\rangle=\frac 14\lim_{x'\rightarrow x}
tr\left[\sigma_\mu(\nabla_\nu-\nabla_{\nu'})+\sigma_\nu(\nabla_\mu-
\nabla_{\mu'})\right]S_F(x,x') \ ,
\end{equation}
by which we have
\begin{equation}
\label{T00}
\langle T_{00}(x)\rangle=i\lim_{x'\rightarrow x}\partial_t^2 
tr\left(G(x,x')\right)=-\lim_{x'\rightarrow x}\partial_\tau^2 
tr\left(G_E(x,x')\right) \ .
\end{equation} 

It was pointed out in \cite{Mello1}, that only the time derivative in the 
bispinor gives a nonzero contribution to the zero-zero component of the 
energy-momentum tensor. Moreover taking the coincidence limit 
$\Omega=\Omega'$ into (\ref{GreenE}), summing over $m_j$ and  using the 
integral representation for the Legendre function \cite{G}, it is possible 
to develop the sum over $j$. Finally we arrive at the following expression 
for the Euclidean bispinor:
\begin{equation}
\label{GE}
G_E(\Delta\tau,r,r')=\frac 1{16\pi^2rr'}\int_b^\infty \frac {dx}
{\sqrt{x^2-b^2}} \frac1{\sinh^2\left(\frac{\arcsinh(x)}
{\alpha}\right)}I_{(2)} \ ,
\end{equation} 
where the function $b$ is expressed in terms of the one-half of the square
of the geodesic distance $\sigma$ in $(\tau,\ r)$ surface
\begin{equation}
b^2=\frac{\alpha^2}{2rr'}\sigma 
\end{equation}
with
\begin{equation}
2\sigma=\Delta\tau^2+\frac{\Delta r^2}{\alpha^2} \ .
\end{equation}

The Green function (\ref{GE}) is divergent in the coincidence limit 
$b\rightarrow 0$. So, in order to obtain a finite and well defined result
for (\ref{T00}), we must renormalize it subtracting all its
divergences. There, it was adopted the point-splitting renormalization
procedure and subtracted from the Green function the Hadamard one,
$G_H(x,x')$, which is expressed in terms of the square of 
geodesic distance $\sigma$. The latter presents short distance
behaviors proportional to $1/\sigma$, as a free theory, plus a logarithmic
scale-dependent term proportional to the scalar curvature of the
spacetime $R\ln(\mu^2\sigma)$.

The thermal average of the energy-momentum tensor of a thermal bath of 
massless
fermions in the global monopole spacetime can be computed in similar way 
replacing the ordinary spinor Greens function in (\ref{T_mu,nu}) by its 
thermal correspondent, $S_\beta(x,x')$, given
in (\ref{S_beta}). For the sake of simplicity let us consider in this paper 
the zero-zero component of the thermal average of the energy-momentum 
tensor, which is also given by
\begin{equation}
\label{T00_beta}
\langle T_{00}(x)\rangle_\beta=-\lim_{x'\rightarrow x}\partial_\tau^2 tr 
\left(G_\beta(x,x')\right) \ .
\end{equation}

The above expression gives us a divergent result which comes exclusively 
from the the zero-temperature contribution of the thermal bispinor, 
the component $n=0$ of (\ref{Green}), which we denominate by $G_\infty(x,x')$.
In fact this singularity is consequence of the evaluation of the Legendre 
function at unity. So in order to obtain a finite and well defined result 
for (\ref{T00_beta}) we have to renormalize its zero-temperature part only, 
$\langle T_{00}(x)\rangle_\infty$. In this way the renormalized thermal 
average of the zero-zero component of the energy-momentum tensor can be 
expressed by
\begin{equation}
\label{T_betaR}
\langle T_{00}(x)\rangle_\beta^{Reg.}=\langle T_{00}(x)\rangle_\infty^{Reg.}+
\langle \overline{T}_{00}(x) \rangle_\beta \ .
\end{equation}

Because the first term in the right hand side of the above equation has
been computed before, in this paper we calculate the purely thermal 
correction, $\langle \overline{T}_{00}(x)\rangle_\beta$. Again taking the 
coincidence limit $\Omega=\Omega'$ into (\ref{Green}) and repeating the same 
procedure adopted for the zero-temperature case, we get
\begin{equation}
\langle \overline{T}_{00}(x)\rangle_\beta=-\lim_{\Delta\tau\rightarrow 0}
\partial_\tau^2 tr(\overline{G}_\beta(\Delta\tau,r)) \ ,
\end{equation}
where
\begin{equation}
\label{G_beta}
\overline{G}_\beta(\Delta\tau,r)=\frac 1{16\pi^2r^2}\sum_{n\neq 0}(-1)^n
\int_{b_n}^\infty \frac {dx}{\sqrt{x^2-b_n^2}} 
\frac1{\sinh^2\left(\frac{\arcsinh(x)}{\alpha}\right)}I_{(2)} \ ,
\end{equation}
being $b_n^2=\frac{\alpha^2(\Delta\tau+n\beta)^2}{4r^2}$. Because $b_n>0$,
the purely thermal correction in (\ref{T_betaR}) is finite. Changing the 
variable $x$ by $b_nx$ in (\ref{G_beta}), taking its second time derivative
followed by the coincidence limit, we find:
\begin{eqnarray}
\label{Energy-Density}
\langle \overline{T}_{00}(x)\rangle&=&-\frac1{8\pi^2r^4}\sum_{n=1}^\infty(-1)^n
\int_1^\infty\frac{dzz^2}{\sqrt{z^2-1}}\left[\frac{2\cosh^2(y)+1}
{\sinh^4(y)\left(1+\frac{z^2\alpha^2n^2\beta^2}{4r^2}\right)}\right.
\nonumber\\
&&\left.-\frac{z\alpha^2n\beta}{2r}\frac{\cosh(y)}{\sinh^3(y)}
\frac1{\left(1+\frac{z^2\alpha^2n^2\beta^2}{4r^2}\right)^{3/2}}\right] \ ,
\end{eqnarray}
with
\begin{equation}
\label{Y}
y=\frac{\arcsinh\left(\frac{z\alpha n\beta}{2r}\right)}\alpha \ .
\end{equation}

Unfortunately, from the above expression, it is not possible to obtain
any information about the behavior of the energy density associated with
a thermal bath of massless fermions in this spacetime as a function of 
the temperature. At this stage we only can observe  that the integrand
decreases exponentially for large values of $z$ and that the series in $n$
converge faster than $1/n^2$ when $n$ goes to infinity. Because we 
want to have a concrete information about the dependence of $\langle T_{00}
\rangle_\beta$ with the temperature, our next steps is to present this
dependence for some specific values of the parameter $\alpha$. In what
follows we present this information for three different situations:
$a)$ $\alpha$ close to unity, small solid angle deficit (excess),
$b)$ small value of $\alpha$, large solid angle deficit and $c)$
large value of $\alpha$, large solid angle excess. (In the Appendix we 
also present some concrete results for $\alpha=1/2$ and brief discussion 
for the cases $\alpha=1/2$ and $\alpha=1/4$.)

\subsection{Parameter $\alpha$ close to the unity.}

In this subsection we present the explicit dependence of the purely
thermal correction to $\langle \overline{T}_{00}(x)\rangle_\beta$ admitting 
that $|1-\alpha|<<1$. In fact, because the global monopole is a very heavy
object which appears as consequence of a breakdown global $O(3)$ gauge
symmetry to $U(1)$, for a typical grand unified theory $1-\alpha^2
\sim 10^{-5}$. So in this case we can expand  (\ref{Energy-Density}) 
in powers of the parameter $\eta^2=1-\alpha^2$. After long calculation
we arrive, up to the first order in $\eta^2$, at:
\begin{eqnarray}
\label{A1}
\langle \overline{T}_{00}(x)\rangle_\beta&=&\frac{7\pi^2}{120\beta^4}
(1+2\eta^2)+\frac{\eta^2}{4\pi^2r^4}\sum_{n=1}^{\infty}
\frac{(-1)^n}{n^2}\frac{\arctanh{\left(\frac1{\sqrt{1+n^2\beta^2/{4r^2}}}
\right)}}{(\beta/r)^2\sqrt{1+n^2\beta^2/{4r^2}}}
\nonumber\\
&&+\frac{2\eta^2}{\pi^2 r^4}\sum_{n=1}^{\infty}\frac{(-1)^n}{n^5}
\frac1{(\beta/r)^5}\int_1^{\infty}\frac{dz}{\sqrt{z^2-1}}\frac1{z^3}
\arcsinh{\left(zn\beta/2r\right)}\times
\nonumber\\
&&\frac{\left[6\left(\frac{zn\beta}
{2r}\right)^4+19\left(\frac{zn\beta}{2r}\right)^2+12\right]}
{\left(1+z^2n^2\beta^2/{4r^2}\right)^{3/2}} \ .
\end{eqnarray}

Again this result also does not clarify completely our objective 
because of the presence of two contributions which cannot be expressed
in terms any functions found in literature. However the above expression has 
some advantage when compared with the previous one obtained without 
approximation, Eqs. (\ref{Energy-Density}) and (\ref{Y}). From (\ref{A1}) 
it is possible to proceed a numerical evaluation of both contributions for 
some specific values of the ratio $\xi:=\beta/r$ and in this way to provide 
a quantitative information for $\langle\overline{T}_{00}\rangle_\beta$. 
Our numerical analysis were developed in the high temperature 
or large distance limit, $\xi<<1$, and the results are shown in Figs. $1(a)$ 
and $1(b)$ for $S_1(\xi)$ and $S_2(\xi)$, representing, respectively, the 
first and second summation as shown below:
\begin{equation}
S_1(\xi)=\sum_{n=1}^{\infty}\frac{(-1)^n}{n^2}\frac1{\xi^2}
\frac{\arctanh{\left(\frac1{\sqrt{1+n^2\xi^2/4}}\right)}}
{\sqrt{1+n^2\xi^2/4}} 
\end{equation}
and
\begin{eqnarray}
S_2(\xi)&=&\sum_{n=1}^{\infty}\frac{(-1)^n}{n^5}\frac1{\xi^5}
\int_1^{\infty}\frac{dz}{\sqrt{z^2-1}}\frac1{z^3}
\arcsinh{\left({zn}\xi/2\right)}\times
\nonumber\\
&&\frac{\left[6\left(\frac{z n\xi}{2}\right)^4+19\left(\frac{zn\xi}{2}\right)^2+12\right]}
{\left(1+z^2n^2\xi^2/{4}\right)^{3/2}} \ .
\end{eqnarray}
From the graphs displayed in Fig. $2$, which exhibits the logarithmic
behavior of the previous summations, it is possible to infer the following
dependence for $S_1$ and $S_2$ with $\xi$:
\begin{equation}
S_1(\xi)=\frac{a_1}{\xi^{q_1}}
\end{equation}
and
\begin{equation}
S_2(\xi)=\frac{a_2}{\xi^{q_2}} \,
\end{equation}
where\, $a_1=-4.701$\, with\, $q_1=2.228\pm0.029$\, and\, $a_2=-8.735$\,  
with\, $q_2=4.000\pm0.000$. From these two results we can see that the most
relevant contribution comes from $S_2$, which, according to the
precision of our numerical analysis, provides to (\ref{A1}) a term
proportional to $1/\beta^4$, so independent of the distance from the
point to the global monopole.

\subsection{Small Parameter: $\alpha<<1$}

In the limit $\alpha<<1$ the leading term in the expansion of the integrand 
of (\ref{Energy-Density}) give us
\begin{equation}
\langle \overline{T}_{00}(x)\rangle_\beta=-\frac1{8\pi^2r^4}
\sum_{n=1}^{\infty}(-1)^n\int_1^{\infty}dz\frac{z^2}{\sqrt{z^2-1}}
\frac{2\cosh^2(zn\beta/2r)+1}{\sinh^4\left(zn\beta/2r \right)} \ ,
\end{equation}
which becomes independent on the parameter $\alpha$. Once more this result 
does not elucidate our main question, which is to provide the information about
the dependence of the thermal correction $\langle \overline{T}_{00}(x)
\rangle_\beta $ with the temperature. Again only numerical evaluation provides
this information. The summation part, denominate by $S_3$, is clearly
a function of the ratio $\xi=\beta/r$. Again our numerical results in the
high temperature regime, $\xi<<1$,  are  exhibited in Fig. $3(a)$ for $S_3$ 
itself and in Fig. $3(b)$ for its logarithmic. From both it is possible to
conclude that
\begin{equation}
S_3(\xi)=\frac{a_3}{\xi^{q_3}} \ ,
\end{equation}
where $a_3=-9.638$ with $q_3=4.000 \pm 0.000$. This result indicates that 
for large solid angle deficit, the high-temperature contribution
to the energy density is uniform and independent on the parameter $\alpha$,
i.e., $\langle T_{00}\rangle_\beta \sim 1/\beta^4$.

\subsection{Large Parameter: $\alpha>>1$}

Now in the limit $\alpha>>1$, the leading term in the expansion of 
(\ref{Energy-Density}) reads:
\begin{eqnarray}
\label{ED1}
\langle \overline{T}_{00}(x)\rangle_\beta&=&-\frac{\alpha^2}{2\pi^2r^2\beta^2}
\left[3\sum_{n=1}^{\infty}\frac{(-1)^n}{n^2}
\int_1^{\infty}\frac{dz}{\sqrt{z^2-1}}
\frac1{\left[\ln\left(\alpha zn\beta/r\right)\right]^4}\right.
\nonumber\\
&&\left.+\sum_{n=1}^{\infty}\frac{(-1)^n}{n^2}\
\int_1^{\infty}\frac{dz}{\sqrt{z^2-1}}
\frac1{\left[\ln\left(\alpha zn\beta/r\right)\right]^3} \right] \ .
\end{eqnarray}

The above result presents two terms whose their sum cannot be
evaluated analytically. They appear as the first terms of an expansion 
of a well defined expression. Consequently they are valid only in the 
case where the argument of the logarithmic never becomes equal to the 
unity. This means that, from the 
numerical point of view, $\alpha$ should assume values bigger than some 
critical one, which depends on the interval domain of the variable 
$\xi$, in such way the divergence is never attained. In order to
proceed a numerical evaluation of both terms, let us write down both
terms in a convenient form, absorbing the factor $r^2/\beta^2=1/\xi^2$. 
Defining them by $S_4$ and $S_5$ they are given as:
\begin{equation}
S_4(\xi)=\sum_{n=1}^{\infty}\frac{(-1)^n}{n^2}\frac1{\xi^2}
\int_1^{\infty}\frac{dz}{\sqrt{z^2-1}}
\frac1{\left[\ln\left(\alpha zn\xi\right)\right]^4}
\end{equation}
and
\begin{equation}
S_5(\xi)=\sum_{n=1}^{\infty}\frac{(-1)^n}{n^2}\frac1{\xi^2}
\int_1^{\infty}\frac{dz}{\sqrt{z^2-1}}
\frac1{\left[\ln\left(\alpha zn\xi\right)\right]^3} \ .
\end{equation}
In this way there appears the factor $1/r^4$ multiplying both terms
in (\ref{ED1}). Adopting for $\xi=\beta/r$ values in the same 
interval used in the other previous numerical analysis our results 
for the two summations can be expressed as shown below:
\begin{equation}
S_4(\xi)=\frac{a_4}{\xi^{q_4}}
\end{equation}
and
\begin{equation}
S_5(\xi)=\frac{a_5}{\xi^{q_5}} \ ,
\end{equation}
where we exhibit in Tables $1$ and $2$ the exponents and respective coefficients
for different values of $\alpha$ for $S_4$ and $S_5$, respectively. 
Unfortunately we could not find any relationship between these parameters 
with $\alpha$, although we can
notice no significant variations on them, i.e., the exponents, $q_4$ and
$q_5$, do not change appreciably when we change $\alpha$. The dominant 
dependence of $\langle \overline{T}_{00} \rangle_\beta$ with $\alpha$ is 
explicitly given by the power $\alpha^2$ in (\ref{ED1}).

\section{Concluding Remarks}

In this paper we have considered the thermal average of the energy 
density associated with a thermal bath of massless
spin-$1/2$ fermions in the background of a pointlike global monopole.
The reason why we have taken into account massless field is because
in this case it is possible to express the spinor Green function in a 
closed form in terms of special functions. Moreover, we decided to
consider only one helicity state, the left-handed one, associated
with the neutrino's degrees of freedom.
In order to develop this analysis we have derived a general expression 
for the respective Euclidean thermal spinor Green function, $S_\beta(x,x')$. 
Using this Green function we obtained a formal expression for the the 
zero-zero component of the thermal average of the energy-momentum tensor, 
$\langle T_{00}\rangle_\beta$, which contains a zero temperature term plus
purely thermal correction. Because this correction is expressed in terms 
of not solvable integrals, its dependence on the temperature could not be 
evaluated explicitly. So, to get some quantitative information, we decided 
to analyze this term for specific values of the parameter
$\alpha$. In this way three formal expressions were obtained for special 
cases related in last section. Unfortunately our results were expressed in 
terms of infinite series not analytically evaluated; however from them it is
possible to develop a numerical analysis. We developed these analysis 
in a high-temperature regime. Our final results enables us to evaluate 
qualitatively the thermal corrections, 
$\langle \overline{T}_{00}\rangle_\beta$. For the cases $\alpha$ close 
and much smaller than unity, the corrections are dominated by terms of 
order $T^4$, consequently independent of the distance from the point to 
the monopole. For the case $\alpha>>1$, there appears an explicit dependence 
of this correction with the distance.

The obtained results may have some applications in the early cosmology,
where the temperature of the Universe was really high. We can see that
the thermal contributions modify significantly the zero-temperature
energy density in that epoch. 

A few years ago, Linet \cite{Linet} has obtained the Euclidean spinor
thermal Green function for a massless spin-$1/2$ fermionic field in the
spacetime of a global cosmic string. There, it was possible to
express the thermal average of the energy-momentum tensor in a closed
form. Linet has found that the high-temperature dominant contribution for
energy density is proportional to $T^4$ and that the
thermal bath of massless fermions is not perturbed by the 
presence of the straight cosmic string\footnote{The high-temperature dominant
contribution given in the Linet's paper, $\langle\overline{T}_{00}(x)
\rangle=\frac{7\pi^2}{60\beta^4}$, is precisely twice bigger than
the flat one found in our result in (\ref{A1}). The discrepancy
between both results is because we considered in this paper just one
helicity state for the fermionic field.}.   
In the pointlike global monopole spacetime we cannot say 
that. For two specific cases treated here the results for thermal 
energy-densities in the 
high-temperature regime present explicit dependence on the parameter
$\alpha$. Only the case where $\alpha<<1$ does not present this dependence.

The contribution due to the temperature in the renormalized vacuum 
expectation value of the square of the massless scalar field and its 
respective energy density in the global monopole spacetime, were recently 
calculated in \cite{Mello2} using similar procedure adopted in this paper.
There appears a non-vanishing contribution to the energy density, coming 
from the geometry of the spacetime, to the dominant $1/\beta^4$ term.

\renewcommand{\theequation}{\Alph{section}.\arabic{equation}}
\appendix

\section{Thermal-Energy Density for Specific Values of $\alpha$}

Here in this appendix we present the purely thermal correction for the energy
density for the cases where the parameter $\alpha$ assume very specific values.
In this first part we take $\alpha=1/2$. For this case it is possible to
develop the integral which appears in (\ref{G_beta}). It is given by
\begin{equation}
\overline{G}_\beta(\Delta\tau,R)=\frac1{64\pi^2r^2}\sum_{n\neq 0}(-1)^n
\left[\frac1{b_n^2}-\frac1{\sqrt{1+b_n^2}}\arctanh\left(\frac1{\sqrt{1+b_n^2}}
\right)\right] \ ,
\end{equation}
with $b_n=\frac{\Delta\tau+n\beta}{4r}$. From the above expression we get, 
after some calculations, the following result:
\begin{eqnarray}
\langle\overline{T}_{00}(x)\rangle_\beta&=&\frac{7\pi^2}{120\beta^4}-
\frac 1{256\pi^2r^4}-\frac1{192\beta^2r^2}+\frac1{256\pi\beta r^3}
\cossech\left(\frac{4\pi r}\beta\right)
\nonumber\\
&&+\frac3{64\beta^2r^2}\cosh\left(\frac{4\pi r}\beta\right)\cossech^2
\left(\frac{4\pi r}\beta\right)
\nonumber\\
&&+\frac1{128\pi^2r^4}\sum_{n=1}^{\infty}(-1)^n
\frac{\arctanh{\left(\frac1{\sqrt{1+n^2\beta^2/{16r^2}}}\right)}}
{\left(1+n^2\beta^2/{16r^2}\right)^{3/2}}
\nonumber\\
&&-\frac3{256\pi^2r^4}\sum_{n=1}^{\infty}(-1)^n\frac{\arctanh{\left(\frac1
{\sqrt{1+n^2\beta^2/{16r^2}}}\right)}}
{\left(1+n^2\beta^2/{16r^2}\right)^{5/2}} \ .
\end{eqnarray}

In the above expression appears two contributions which cannot
be evaluated analytically. Again if we are inclined to know how these
terms behave as a function of the variable $\xi$ we have to proceed a
numerical investigation. The explicit expressions writing in terms of
$\xi$ are given below:
\begin{equation}
S_6(\xi)=\sum_{n=1}^{\infty}(-1)^n
\frac{\arctanh{\left(\frac1{\sqrt{1+n^2\xi^2/16}}\right)}}
{\left(1+n^2\xi^2/16\right)^{3/2}}
\end{equation}
and
\begin{equation}
S_7(\xi)=\sum_{n=1}^{\infty}(-1)^n\frac{\arctanh{\left(\frac1
{\sqrt{1+n^2\xi^2/16}}\right)}}
{\left(1+n^2\xi^2/16\right)^{5/2}} \ .
\end{equation}
 
From our numerical analysis, developed in the high-temperature regime,
we obtained the following results:
\begin{equation}
S_6(\xi)=\frac{a_6}{\xi^{q_6}}
\end{equation}
and
\begin{equation}
S_7(\xi)=\frac{a_7}{\xi^{q_7}} \ ,
\end{equation}
where $a_6=-0.552$ with $q_6=0.184 \pm 0.019$ and $a_7=-0.552$ with 
$q_7=0.184 \pm 0.019$. So we can conclude that these two terms present
the same order of correction to $\langle \overline{T}_{00} \rangle_\beta$
and are not relevant. So the dominant contribution comes from the term
independent of the distance from the point to the global monopole,
followed by a term of order $1/r^2\beta^2$.

For the case $\alpha=1/3$ the procedure is similar to this previous one.
The integration in (\ref{G_beta}) can be done and after some intermediate 
steps it is possible to conclude that the dominant contribution for
the energy density, in high-temperature limit is also given by
\begin{equation}
\langle\overline{T}_{00}(x)\rangle_\beta=\frac{7\pi^2}{120\beta^4}+
O\left(\frac1{r^2\beta^2}\right) \ .
\end{equation}
The same dominant behavior is obtained for the case $\alpha=1/4$. We shall
not repeat the intermediate steps.

As our final comment in this appendix, we want to say that the dominant 
contributions for $\langle\overline{T}_{00}(x)\rangle_\beta$ in these three 
examples, coincides with the result found for a flat spacetime, 
$\langle\overline{T}_{00}(x)\rangle_\beta=\frac{7\pi^2}{120\beta^4}$. 
However, sub-dominant contributions are consequence of the non flat geometry 
of this spacetime.

\newpage

\section{Figure Captions}

{\bf Figure 1}: The figures 1(a) and 1(b) exhibit, respectively, the 
behavior of $S_1$ and $S_2$ with $\xi$ in the region $[0.01; 0.1]$.\hfill\\
[6mm]
{\bf Figure 2}: This figure exhibits the logarithmic behavior for both 
functions, $S_1$ and $S_2$ with  the logarithm of $\xi$, in the region 
$[0.01; 0.1]$. From it, it is possible to estimate the leading dependence 
for both quantities with $\xi$.\hfill\\[6mm]
{\bf Figure 3}: The figure 3(a) exhibits the behavior of $S_3$ with $\xi$ 
in the region $[0.01; 0.1]$. The figure 3(b) exhibits the logarithmic 
behavior for $S_3$ with the logarithm of $\xi$, in the region $[0.01; 0.1]$. 
From it, it is possible to estimate its leading dependence with $\xi$.\hfill\\
[6mm]
{\bf Table 1}: The table 1, exhibits the exponents $q_4$ with their respective 
coefficients $a_4$ of $S_4$ for different values of $\alpha$.\hfill\\[6mm]
{\bf Table 2}: The table 2, exhibits the exponents $q_5$ with their respective 
coefficients $a_5$ of $S_5$ for different values of $\alpha$.\hfill\\[6mm]

\newpage

\begin{figure}[t]
\begin{center}
\includegraphics[width=6cm]{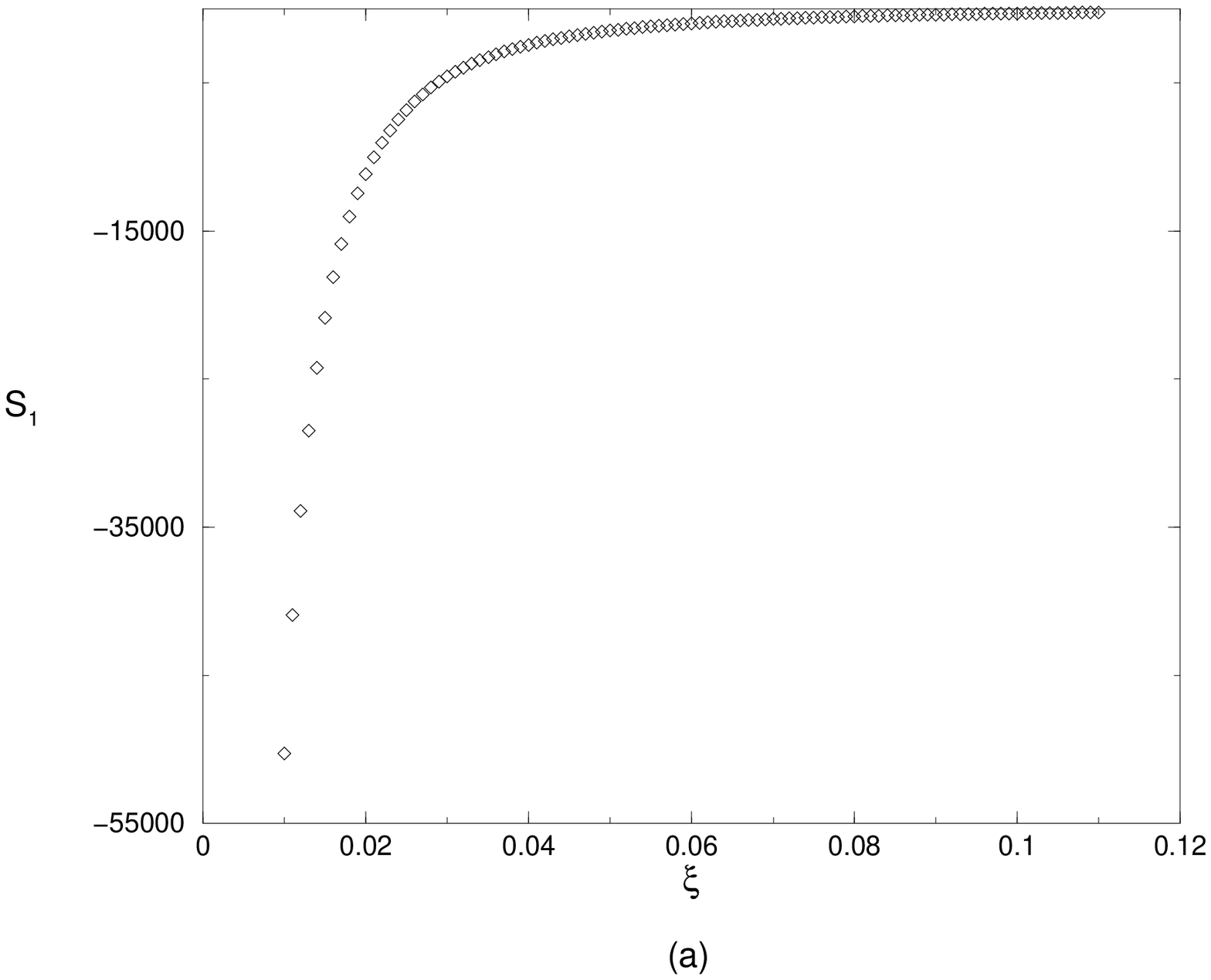}
\includegraphics[width=6cm]{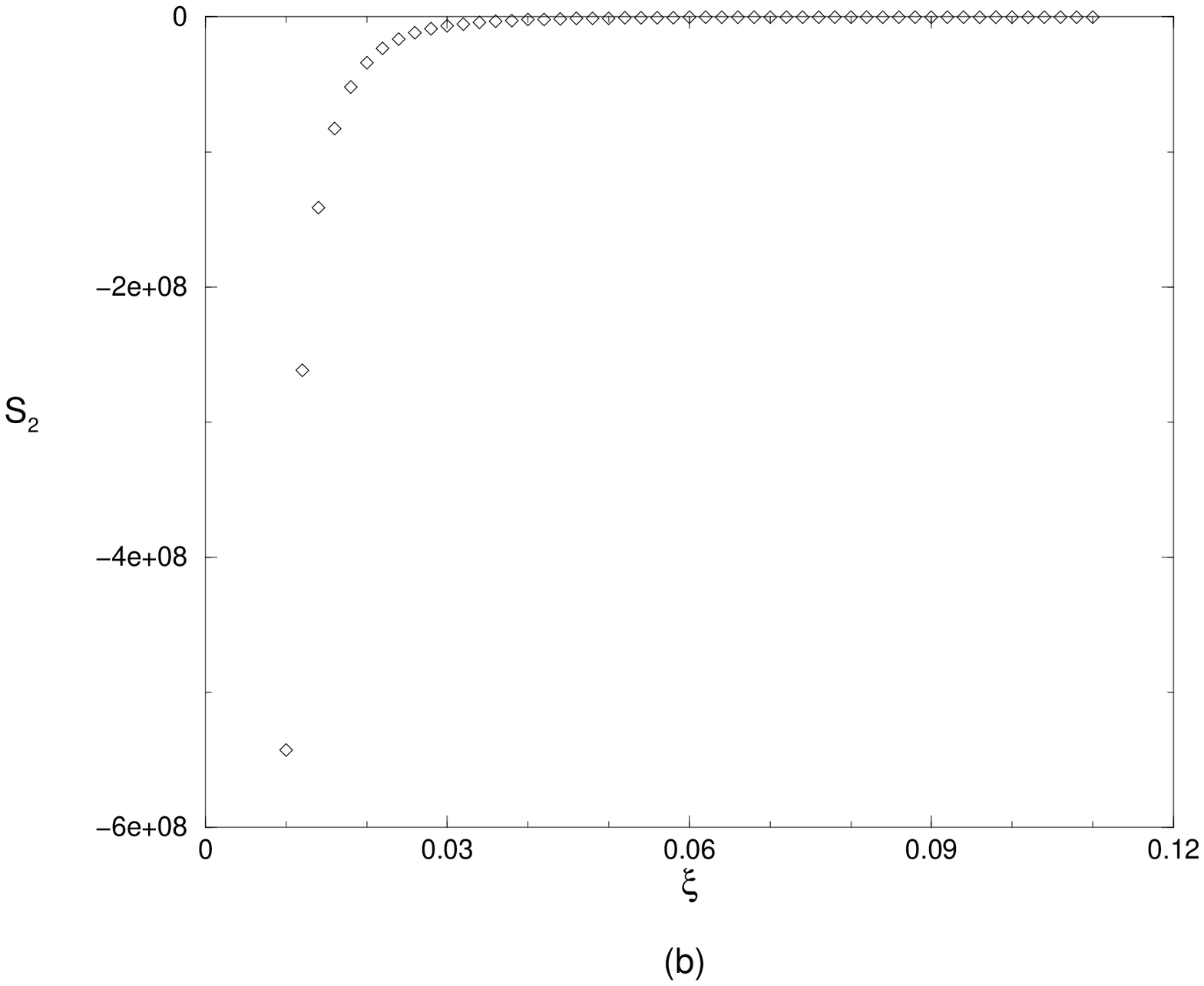}
\label{fig1}
\caption{}
\end{center}
\end{figure}

\begin{figure}[h]
\begin{center}
\includegraphics[width=6cm]{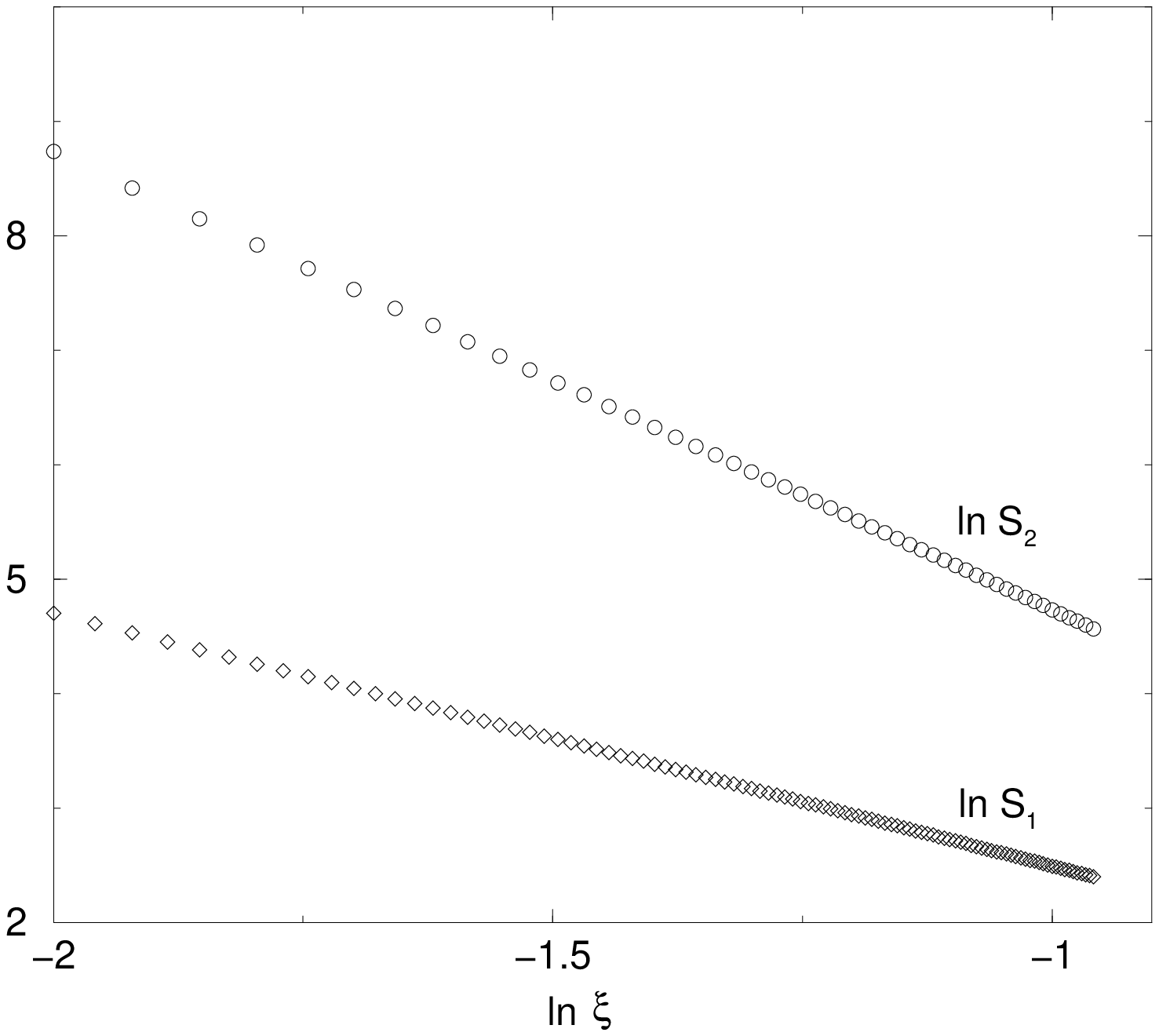}
\label{fig2}
\caption{}
\end{center}
\end{figure}

\begin{figure}[h]
\begin{center}
\includegraphics[width=6.85cm]{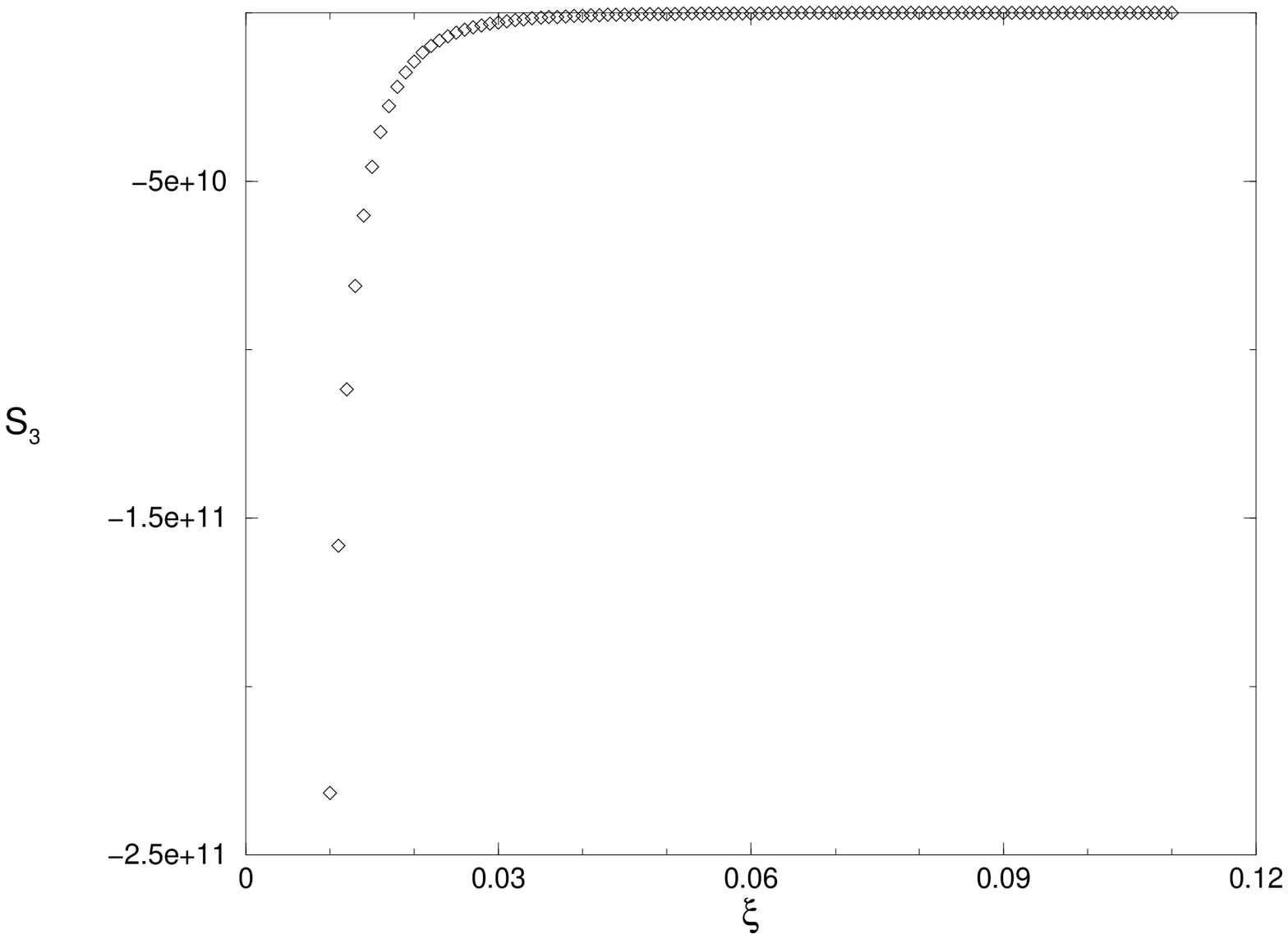}
\includegraphics[width=6cm]{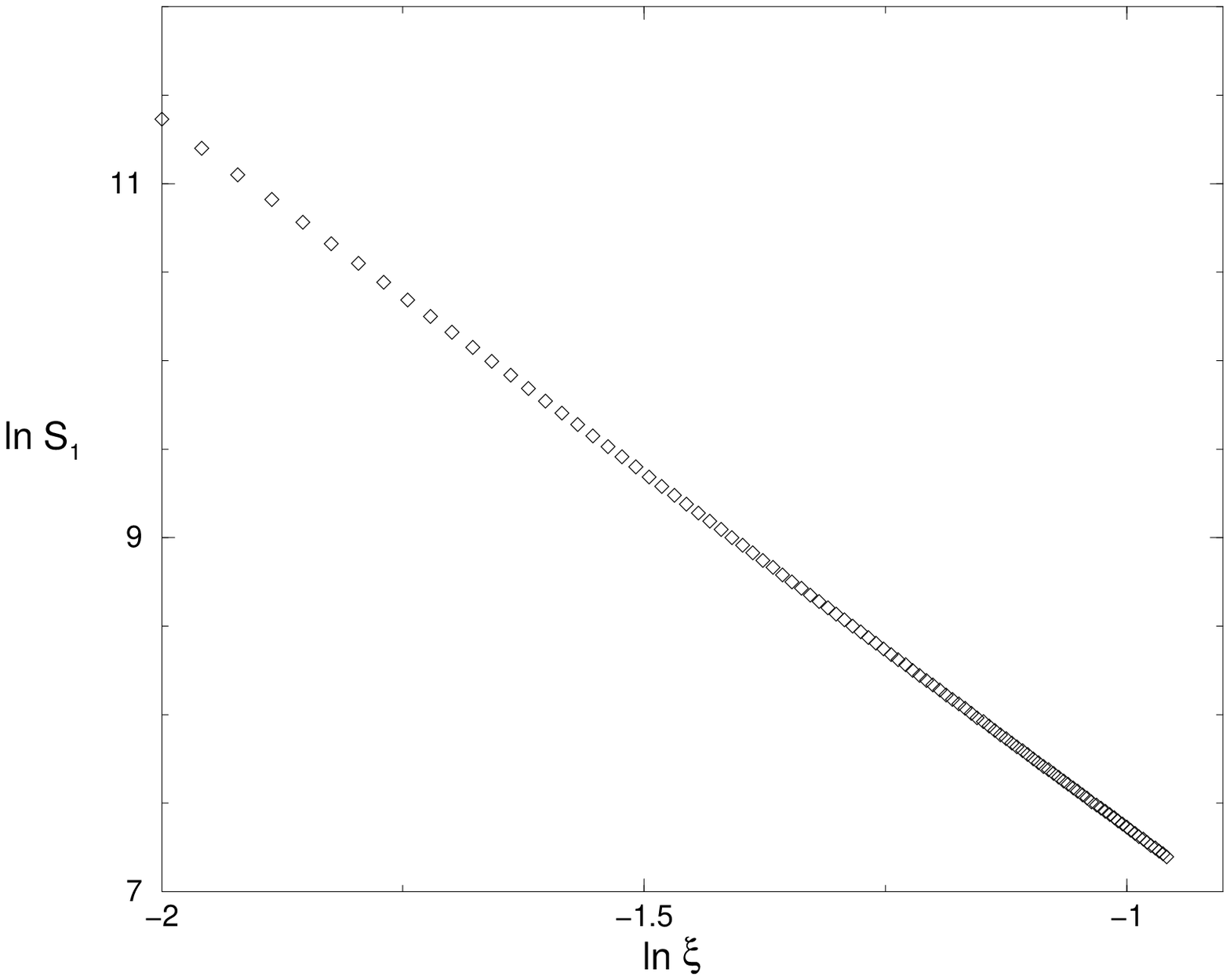}
\caption{}
\label{fig3}
\end{center}
\end{figure}

\begin{table}[p]
\label{tab1}
\begin{center}
\begin{tabular}{||l|l|l||} \hline\hline
\multicolumn{1}{||c}{} & \multicolumn{1}{c}{\,\,\,\,\,\,\,Table 1} & \multicolumn{1}{c||}{}\\ \hline
\multicolumn{1}{||c|}{$\alpha$} & \multicolumn{1}{c|}{$q_4$} & \multicolumn{1}{c||}{$a_4$}\\ \hline
$1000$ & $2.930375608\pm0.1669581160$ & $-2.596702947$\\ \hline
$1500$ & $2.8414692776\pm0.1335029194$ & $-2.348446259$\\ \hline
$2000$ & $2.788466901\pm0.1156906043$ & $-2.193442147$\\ \hline
$2500$ & $2.751914554\pm0.1042923505$ & $-2.082895068$\\ \hline
$3000$ & $2.724562303\pm0.09622178978$ & $-1.997972468$\\ \hline\hline
\end{tabular}
\caption{}
\end{center}
\end{table}

\begin{table}[p]
\label{tab2}
\begin{center}
\begin{tabular}{||l|l|l||} \hline\hline
\multicolumn{1}{||c}{} & \multicolumn{1}{c}{\,\,\,\,\,\,\,Table 2} & \multicolumn{1}{c||}{}\\ \hline
\multicolumn{1}{||c|}{$\alpha$} & \multicolumn{1}{c|}{$q_5$} & \multicolumn{1}{c||}{$a_5$}\\ \hline
$1000$ & $2.632887909\pm0.12000918880$ & $-3.059963397$\\ \hline
$1500$ & $2.569103696\pm0.09581311868$ & $-2.887785837$\\ \hline
$2000$ & $2.531144121\pm0.08291154623$ & $-2.780842812$\\ \hline
$2500$ & $2.505004736\pm0.07464879774$ & $-2.704862735$\\ \hline
$3000$ & $2.485469694\pm0.06879530831$ & $-2.646669348$\\ \hline\hline
\end{tabular}
\caption{}
\end{center}
\end{table}


\newpage
\begin{thebibliography}{100}
\bibitem{Mello1} E. R. Bezerra de Mello, V. B. Bezerra and  
N. R. Khusnutdinov, Phys. Rev. D {\bf 60}, 063506 (1999).
\bibitem{BV} M. Barriola and A. Vilenkin, Phys. Rev. Lett. {\bf 63}, 341
(1989).
\bibitem{Jackiw} L. Dolan L and R. Jackiw, Phys. Rev. D {\bf 9}, 3320 
(1974).
\bibitem{BD} J. D. Bjorken and S. D. Drell, {\it Relativistic Quantum 
Mechanics}
(McGraw-Hill, New York, 1964).
\bibitem{G}  I. S. Gradshteyn and I. M. Ryzhik, {\it Table of Integrals, 
Series and Products}, Academic Press, Inc. 1980 .
\bibitem{Braden}H. W. Braden, Phys. Rev. D {\bf 25}, 1028 (1982).
\bibitem{Birrel} N. D. Birrell and P. C. W. Davis, {\it Quantum Fields in 
Curved Space} (Cambridge University Press, Cambridge, England, 1982).
\bibitem{Wald} R. M. Wald, Commun. Math. Phys. {\bf 54}, 1 (1977).
\bibitem{CD} S. M. Christensen and M. J. Duff, Phys. Lett. {\bf76B}, 
571 (1986).
\bibitem{Christensen} S. M. Christensen, Phys. Rev. D {\bf 17}, 946 (1978).
\bibitem{Linet} B. Linet, Class. and Quantum Gravity, {\bf 13}, 97 (1996).
\bibitem{Mello2} F. C. Cabral and E. R. Bezerra de Mello, Class.
Quantum Grav. {\bf 18}, 1637 (2001).
\end{thebibliography}
\end{document}